\documentclass[a4paper]{jpconf}
\usepackage{graphicx,url}
\usepackage{amsmath,amssymb}
\usepackage{color}

\usepackage{ulem}   

\renewcommand{\vec}[1]{\boldsymbol{#1}}

\begin{document}

\title{Detection of GW bursts with chirplet-like template families}
\author{\'Eric Chassande-Mottin$^{\star}$, Miriam Miele$^{\star \dagger}$, Satya Mohapatra$^{\ddagger}$ and Laura Cadonati$^{\ddagger}$ } 
 \address{$^{\star}$ CNRS and Univ. Paris Denis Diderot, AstroParticule et Cosmologie (France)\\
$^{\dagger}$ ERASMUS fellow from the University of Sannio at Benevento (Italy)\\
$^{\ddagger}$  Physics Department, University of Massachusetts, Amherst MA 01003  (USA)}

\ead{ecm@apc.univ-paris7.fr}

\begin{abstract}
  Gravitational Wave (GW) burst detection algorithms typically rely on the hypothesis that the
  burst signal is ``locally stationary'', that is  it changes slowly with frequency.
  Under this assumption, the signal can be
  decomposed into a small number of wavelets with constant frequency.  This
  justifies the use of a family of sine-Gaussian wavelets in the Omega
  pipeline, one of the algorithms used in LIGO-Virgo burst searches. 
  However there are plausible scenarios where the burst
  frequency evolves rapidly, such as in the merger phase of a binary black hole and/or
  neutron star coalescence.  In those cases,
  the local stationarity of sine-Gaussians induces performance losses, due to
  the mismatch between the template and the actual signal. We propose an
  extension of the Omega pipeline based on chirplet-like templates. Chirplets
  incorporate an additional parameter, the \textit{chirp rate}, to control
  the frequency variation. In this paper, we show that the Omega pipeline can
  easily be extended to include a chirplet template bank. We illustrate the
  method on a simulated data set, with a family of 
  phenomenological  binary black-hole coalescence waveforms
  embedded into Gaussian LIGO/Virgo--like noise. Chirplet-like templates result
  in an enhancement of the measured signal-to-noise ratio.
 \end{abstract}

\section{Motivations}

Current searches for gravitational wave transients in LIGO-Virgo data focus on two signal classes:
short unmodelled bursts and longer quasi-periodic signals from
inspiralling black hole and/or neutron star binaries as predicted by
post-Newtonian approximations. 
To account for intermediate scenarios, we 
 consider  ``chirping burst'' GW target signals that exhibit characteristics from both the
above categories: a short duration and a ``sweeping'' frequency.

We propose here an extension of the Omega pipeline \cite{chatterji05:_ligo}
(originally known as $Q-$pipeline) that searches for chirping bursts. The
Omega pipeline projects the data over a family of sine-Gaussian wavelets with
fixed frequency. The idea is to replace these templates by frequency varying
waveforms, referred to as \textit{chirplets}.

In this paper, we first define chirplets and the related \textit{chirplet
  transform}. We discuss the implementation of the chirplet transform and its
insertion into the Omega pipeline, with attention to how the chirplet template bank is built.
Finally, we present a few  examples using simulated data.

\section{From wavelets to chirplets}

\subsection{Definition of chirplets}

\begin{figure}
\centerline{\includegraphics[width=.9\textwidth]{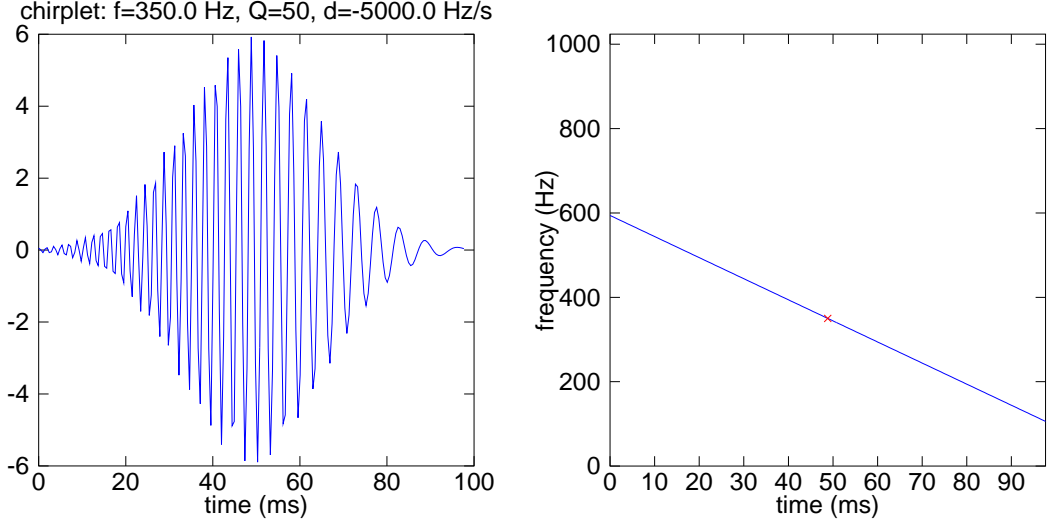}}
\caption{\label{chirplet}Example of a chirplet}
\end{figure}

Chirplets are defined in the time domain as:
\begin{equation}
\psi(\tau)\equiv A \exp \left( -\frac{(2\pi f)^2}{Q^2} (\tau-t)^2\right)
\exp \left( 2\pi i \left[ f (\tau-t) + d/2\: (\tau-t)^2 \right]\right),
\end{equation}
with $A=(8\pi f^2/Q^2)^{1/4}$, a normalization factor ensuring that $\int
|\psi|^2=1$. $t$ and $f$ are the center time and frequency, respectively and $Q$
is the dimensionless quality factor. See in Fig. \ref{chirplet} for an example
of a chirplet.

The main difference from a  sine-Gaussian wavelets is the \textit{chirp rate}, an
additional term in the phase denoted $d$ that changes the chirplet frequency 
linearly in time as $f(\tau)=f+d (\tau-t)$. The chirp rate controls the slope of
the frequency evolution. When $d=0$, we retrieve standard sine-Gaussians.
Chirplets are thus associated with a four-dimensional parameter space instead of
three for sine-Gaussians. In the sequel, we will concatenate all those parameters
into a descriptor $\vec{\theta}\equiv \{t, f, Q, d\}$.

\subsection{Chirplet transform}

The \textit{chirplet transform} $T$ is obtained by correlating the data with the
chirplets defined in the previous section. In the frequency domain, it reads:
\begin{equation}
  \label{ct}
  T[x;\vec{\theta}]= \left| \int X(\xi) \Psi^*(\xi;\vec{\theta}) d\xi \right|^2,
\end{equation}
where $X(\cdot)$ and $\Psi(\cdot;\vec{\theta})$ denotes the Fourier transform of
the (whitened) data stream $x(\cdot)$ and chirplet $\psi(\cdot)$ with descriptor
$\vec{\theta}$ resp.

The chirplet Fourier transform can be expressed as
\begin{equation}
  \label{fourier}
  \Psi(\xi;\vec{\theta})={\cal A} \exp \left(- \frac{\tilde{Q}^2}{4} \frac{(\xi-f)^2}{f^2} \right),
\end{equation}
where ${\cal A}=[(\tilde{Q}^4/Q^2)/(2\pi f^2)]^{1/4}$ is
written in terms of a ``complex-valued'' quality factor $\tilde{Q}= Q
\sqrt{z}/|z|$ where $z=1+i d \Delta_t^2$ with the
chirplet duration\footnote{By definition, $\Delta_t \equiv 2\sqrt{\pi} \int (\tau-t)^2 \psi^2(\tau) d\tau$.} 
$\Delta_t= Q/\left(2\sqrt{\pi} f\right)$.


\section{Building template banks with chirplets}
\label{sec:templatebank}

By varying the chirplet parameters, we obtain a continuous signal space. In this
space, we need to select a finite-size family of representatives which will be
used to analyze the data. The coverage of the chirplet space has to meet two
conflicting goals i.e., satisfy a worst case mismatch with a minimum number of
templates. We adopt the method proposed in
\cite{owen99:_match,chatterji05:_ligo} which consists of sampling the space with
equi-spaced templates using the intrinsic \textit{metric} deduced from the
\textit{mismatch} $T\left[\psi(\vec{\theta}^{\prime});\vec{\theta}\right]$
between two neighboring templates with a small discrepancy
$\delta\vec{\theta}\equiv \vec{\theta}^{\prime}-\vec{\theta}$ in their
parameters. The metric results from the second-order expansion of the mismatch
$\mu_{\vec{\theta}}(\delta \vec{\theta}) \equiv
1-T\left[\psi(\vec{\theta}+\delta \vec{\theta});\vec{\theta}\right] \approx
\delta s^2$ when $\delta\vec{\theta} \rightarrow \vec{0}$ and leads
to\footnote{This calculation assumes that the detector noise has a flat spectrum. Contrarily to
  the sine-Gaussian case, this approximation has significant effect since the
  chirplet frequency varies across the detector bandwidth.}:
\begin{equation}
\label{metric}
\delta s^2 = \frac{Q^4 d^2 + 16 \pi^2 f^4}{4 Q^2 f^2} \delta t^2 +
\frac{2+Q^2}{4 f^2} \delta f^2 + \frac{\delta Q^2}{2 Q^2}
+ \frac{Q^4}{128 \pi^2 f^4} \delta d^2 - \frac{Q^2 d}{2 f^2} \delta t
\delta f - \frac{\delta f \delta Q}{Q f}.
\end{equation}

There are several differences and additional terms from the sine-Gaussian metric,
due to the non-zero chirping rate.  Along the time
axis and for small $f \lesssim Q\sqrt{d}$, the sampling step $\delta t \propto
f/(Qd)$ is finer than that of the sine-Gaussian case $\delta t \propto Q/f$.  We
note also that the sampling step along the chirping rate axis scales with
$\delta d \propto (f/Q)^2$. We thus expect to get many chirplets in the
low-frequency band and for large values of $Q$.

\begin{figure}
\centerline{\includegraphics[width=.7\textwidth]{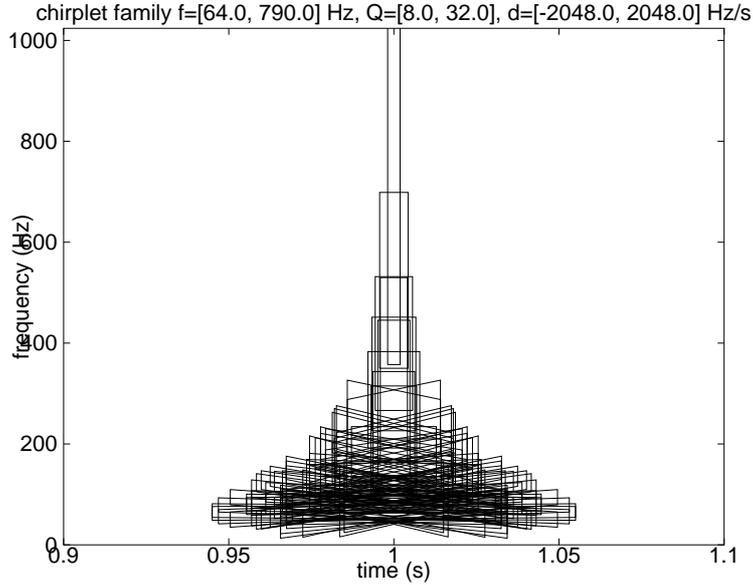}}
\caption{\label{chirpletfamily} Example of a chirplet family resulting from the
  template placement procedure presented in Sec. \ref{sec:templatebank}. In this
  graph, each box represents the time-frequency tile associated with a
  chirplet. The oblique boxes are associated to non-zero values of the chirping
  rate $d$. The slope of the upper and lower edges equals $d$.}
\end{figure}

The chirplet space equipped with the above metric (off-diagonal terms being
neglected for simplicity \cite{chatterji05:_ligo}) can be discretized by a cubic lattice with templates placed at the
vertices. The worst case occurs when the real signal is farther apart from all
vertices, at the center of the cube. Let us denote $\delta s=\mu^{1/2}$, this
worst-case distance, which corresponds to the half-length of the cube diagonal
and assign a maximum value $\mu^{1/2}_{\mathrm{max}}$ that we can
tolerate. Since we are in a four-dimensional space, the length $\ell$ of the
cube edge is equal to that of its half-diagonal. Therefore, we must have $\ell
\leq \sqrt{\mu_{\mathrm{max}}}$.  The discretization along each axis of the
parameter space which results directly from this condition ensures that
$\mu_{\vec{\theta}}(\vec{\theta}-\vec{\theta}_n) \leq \mu_{\mathrm{max}}$ for
any $\vec{\theta}$ with $\vec{\theta}_n$ the closest vertex of the lattice.
In the following, we set the maximum mismatch to 
the value $\mu_{\mathrm{max}}=0.2$ typically used when applying standard Omega.
Fig. \ref{chirpletfamily} shows an example of a chirplet template bank resulting from
this template placement scheme.

\begin{figure}
\centerline{\includegraphics[width=.45\textwidth]{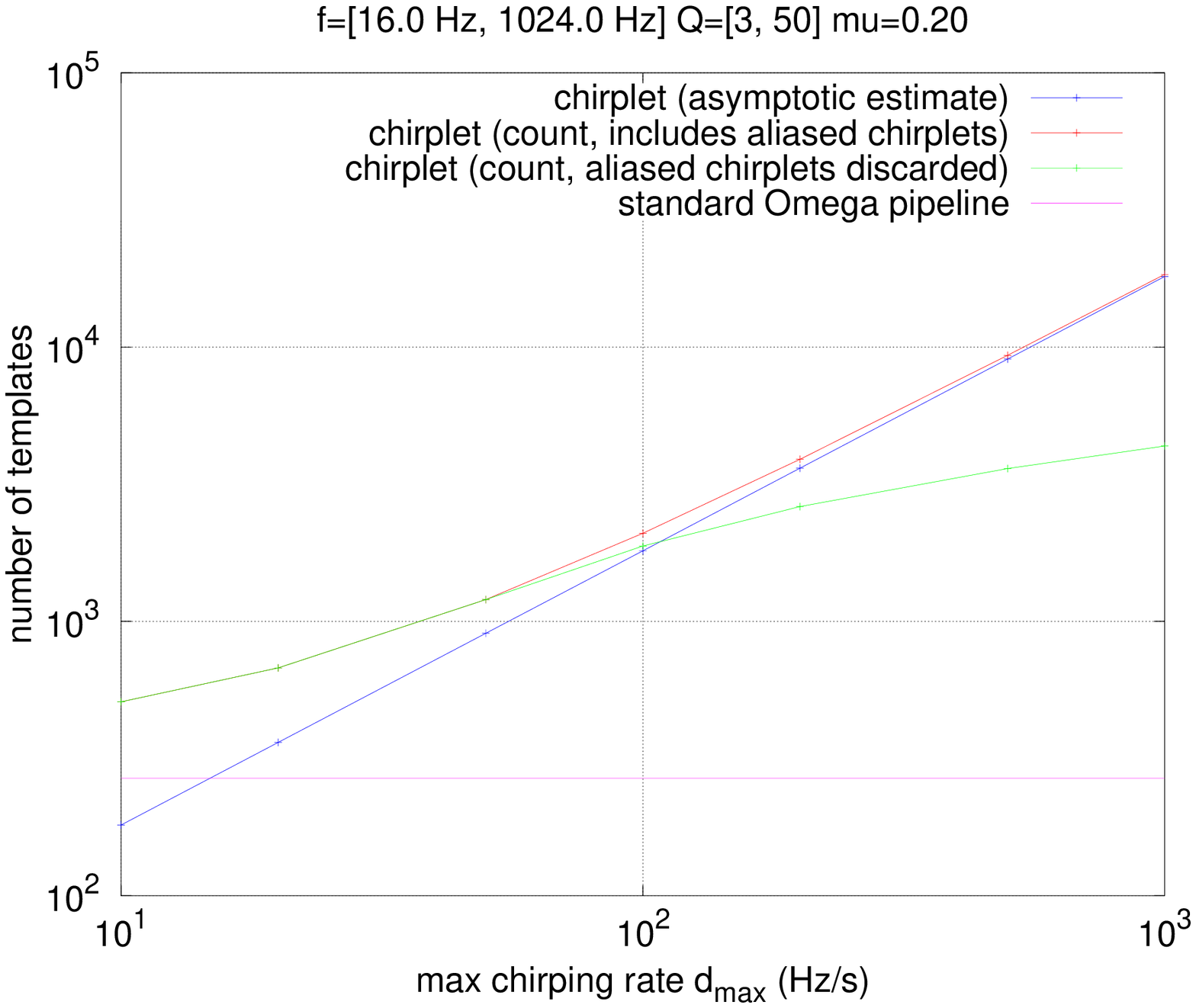}
\includegraphics[width=.45\textwidth]{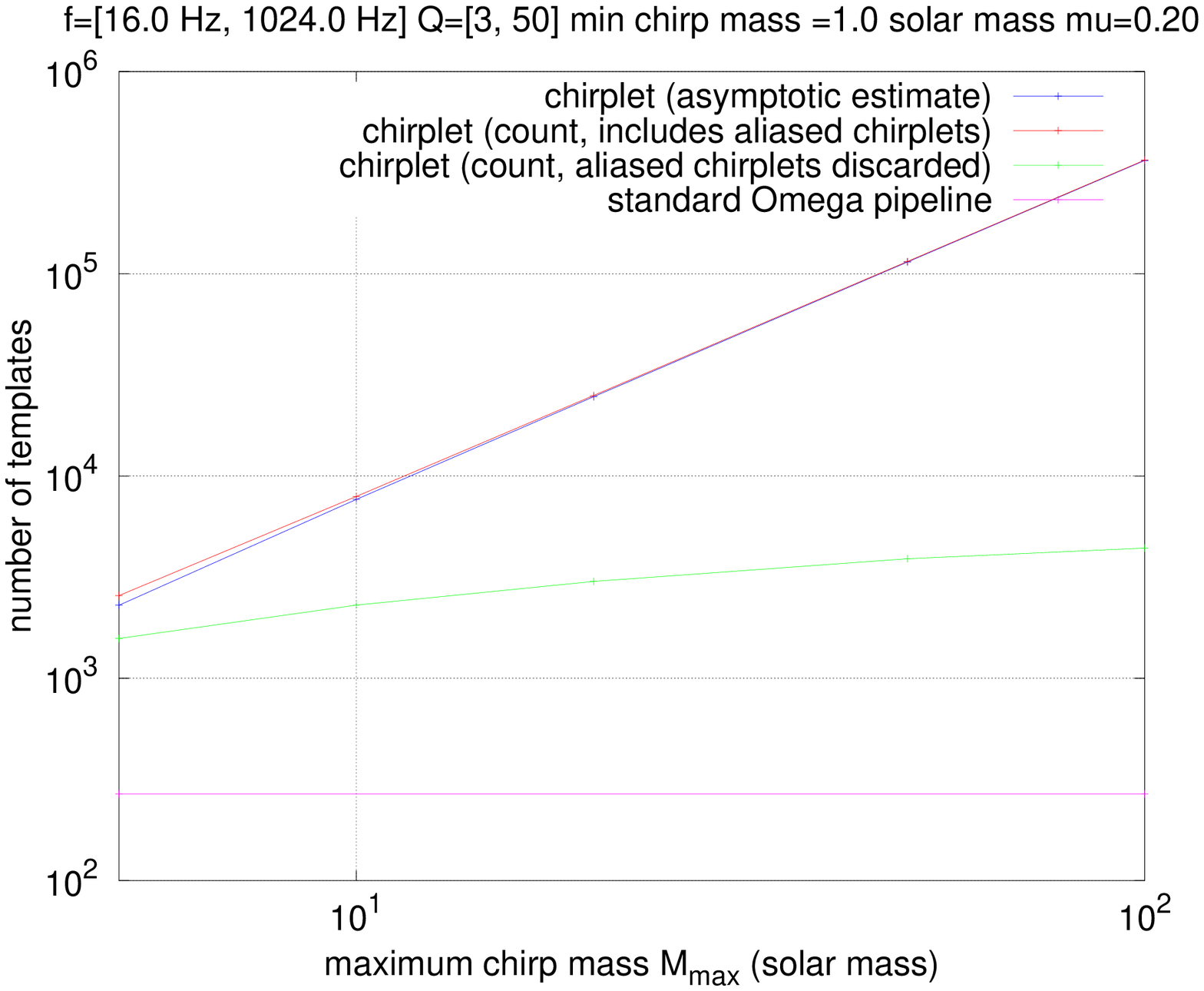}}
\caption{\label{number_template}Size of the chirplet template bank in two cases:
  (\textit{left}) assuming chirp rate limits between $\pm d_{\mathrm{max}}$ for
  any frequencies; (\textit{right}) assuming frequency dependent limits
  consistent with the Newtonian model of the inspiralling binary chirp: $C
  M_{\mathrm{min}}^{5/3} f^{11/3} \lesssim d \lesssim C M_{\mathrm{max}}^{5/3}
  f^{11/3}$.  The results are compared to the size of the sine-Gaussian template
  bank used by standard Omega. See text for a detailed discussion. All template
  banks are generated using the same maximum mismatch $\mu_{\mathrm{max}}=0.2$.}
\end{figure}

In Fig. \ref{number_template}, we apply the same scheme in two different
settings. In both cases we computed the number of templates necessary to cover
the signal space in the sine-Gaussian (standard Omega) and chirplet
(chirpletized Omega) cases. This computation is done at a fixed time $t$.  We
compare the result to the estimate given by the ratio of the whole space volume
$V=\int |\vec{\mu}^{\star}|^{1/2} d^3\vec{\theta}^{\star}$ (where
$\vec{\theta}^{\star}=\{f, Q, d\}$ and $\delta s^2=|\vec{\mu}^{\star}|$ denotes
the metric in Eq. (\ref{metric}) without the components associated to the time
axis) to the size of a cubic element of the lattice. We find\footnote{This
  result is valid both when neglecting or retaining the off-diagonal terms of
  the metric. The scaling is actually exact when the off-diagonal terms are
  included and valid to a good approximation when $Q_{\mathrm{max}} \gg 2$ in
  the other case.}:
\begin{equation}
\label{templatenumber}
{\cal N}\equiv V/\ell^3 \propto f_{\mathrm{min}}^{-2} Q_{\mathrm{max}}^3 d_{\mathrm{max}},
\end{equation}
where we assume that for each coordinate the lower boundaries  (\emph{min}) are
much smaller than the higher boundaries (\emph{max}).

It is important to note that both the count and estimate are obtained assuming
an infinite bandwidth. Since the data are sampled, we are restricted to a
limited Nyquist frequency. Chirplets with frequency exceeding this limits are
aliased and have to be discarded. Fig. \ref{number_template} also show the number
of non-aliased chirplets. This number is about a factor of 10 larger than the
number of sine-Gaussians required to cover the entire parameter space.

Two comments can be made at this point: the larger number of templates indicate
that with chirplets, we define and explore a much larger signal space than with
sine-Gaussians (we investigate this question further in Sec. \ref{sec:action}).
As the computing cost scales approximately linearly with the number of
templates, analyzing the data with chirplets requires with a ten-fold increase
in computing resources (a factor that will be rapidly absorbed by the
exponential growth of computing power).

\section{From ``standard'' to ``chirpletized'' Omega}

In this session we discuss other aspects of the analysis pipeline, in addition to the
implementation of the chirplet template bank.

\subsection{Filtering}

The modulus of $\Psi(\cdot)$ in Eq. (\ref{fourier}) is a Gaussian function as
in the sine-Gaussian case. This allows the use of the same filtering
scheme as in the standard Omega pipeline to generate the chirplet transform.
The Omega  scheme \cite{chatterji05:_ligo} operates in the frequency
domain following Eq. (\ref{ct}). It consists in multiplying the Fourier
transform of the data, computed with the FFT algorithm, with that of the
templates and take the inverse Fourier transform 
of the product. Omega uses a bi-square frequency window that approximates the Gaussian
shape. The compact support of the bi-square window prevents aliasing.

This scheme can be applied to the chirplet case with two differences. First, the
template $\Psi$ is now complex, thus we need to multiply the data spectrum both
in modulus and phase. Second, the template bandwidth now results from the
quadratic sum $\Delta_f^2 =
(\Delta_f^{\mathrm{finite~size}})^2+(\Delta_f^{\mathrm{chirp}})^2$ of two
components, one due to the finite size of the chirplet
$\Delta_f^{\mathrm{finite~size}}=1/\Delta_t$ and the other due to its sweeping
frequency $\Delta_f^{\mathrm{chirp}}=d \Delta_t$, where $\Delta_t$ is the
chirplet duration (defined above). According to \cite{chatterji05:_ligo}, the
width of the bi-square window should be set to the chirplet frequency bandwidth
$\Delta_f$ rescaled by a factor of $\sqrt{11}$.

\subsection{Pre- and post-processing}

In this paper we focus on a single-detector network, where most of the pre- and
post-processing can be adopted from the standard Omega pipeline.  
Pre-processing   consists of whitening the input data
stream.  Post-processing consists of selecting among the chirplets with
partial time and frequency overlap to the one with maximum
correlation with the data.  Each chirplet is associated with a time-frequency tile,
 defined by $[t\pm \Delta_t/2, f \pm \Delta_f/2]$ where
$\Delta_t$ and $\Delta_f$ are the chirplet duration and bandwidth, respectively.
Two chirplets overlap if their time-frequency tiles overlap.

\section{Performances of Chirpletized Omega}
\label{sec:action}

In this section we present a comparison between the standard version of the
Omega pipeline, which uses sine-Gaussian wavelets, and its \emph{chirpletized}
version.  We configure the pipelines with identical values for the parameters
they have in common (frequency and Q range and maximum mismatch). We identify
cases where we can expect advantages from analyzing the data with chirplets.

\subsection{Analyzing chirplets with sine-Gaussians}

\begin{figure}
  \centerline{\includegraphics[width=.7\textwidth]{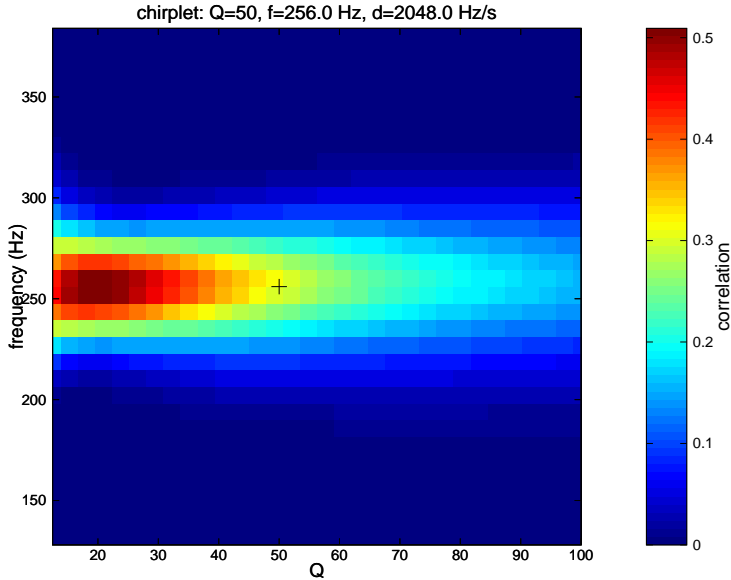}}
  \caption{\label{density} Correlation measurement between a single chirplet and
    a sine-Gaussian template bank. The parameters of the analyzed chirplet is
    indicated with a black cross in this diagram.}
\end{figure}

The signal space associated with sine-Gaussians is contained in the larger space associated
with chirplets. We estimate the signal-to-noise ratio (SNR) loss
occurring when analyzing a chirplet by correlating this signal against a
sine-Gaussian template bank. The chirplet parameters have been set to $Q=50$,
$f=256$ Hz and $d=2048$ Hz/s. Those parameters correspond to observable physical
signals in the LIGO/Virgo frequency band (for instance, the selected chirping
rate is approximately that of an inspiralling binary chirp with total mass $M
\sim 3 M_{\odot}$ -- assuming equal masses -- at $f=256$ Hz according to the
Newtonian model).  Fig. \ref{density} presents the result of this analysis.
Consistently to the metric estimate, the loss is $\sqrt{128}\pi f^2/(dQ^2) \sim
50 \%$ in the present case. Note also that the maximum correlation is shifted to
lower $Q$ which may lead to a possible bias in the estimation of this parameter.

\subsection{Analyzing inspiralling binary chirps with chirplets}

\paragraph{One case study ---}

As an illustration, we show here results from chirpletized Omega on simulated
Gaussian noise, colored with the spectral characteristics of LIGO/Virgo noise
with simulated gravitational wave signals. The signal we consider here results
from the phenomenological approximation introduced in~\cite{ajith-2009} of the
coalescing binary black-hole chirp signals and includes the inspiral, merger and
ringdown parts of the coalescence.

\begin{figure}
\begin{tabular}{cc}
\includegraphics[width=.45\textwidth]{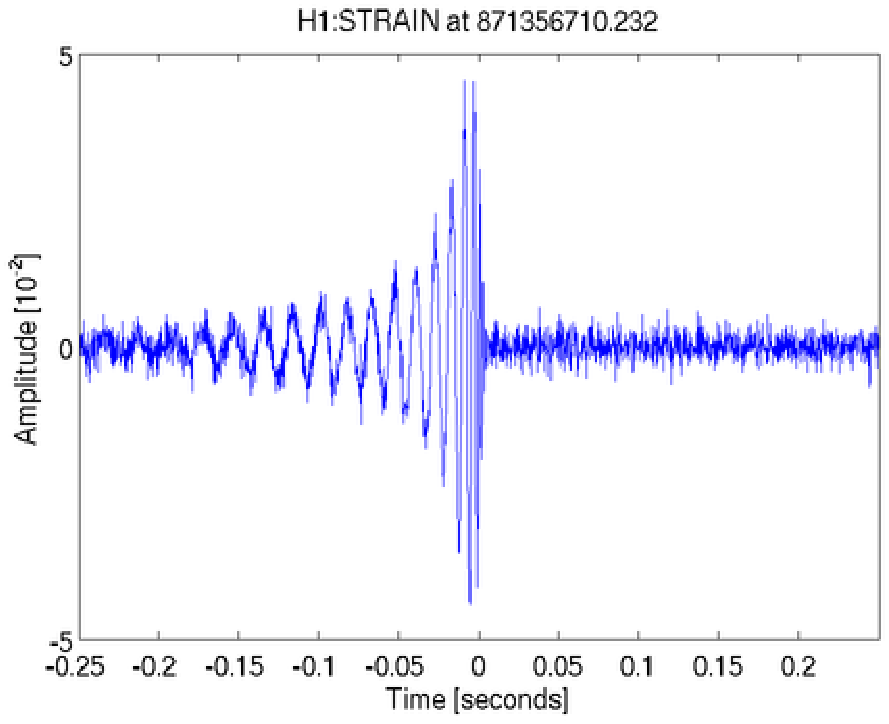}&
\includegraphics[width=.45\textwidth]{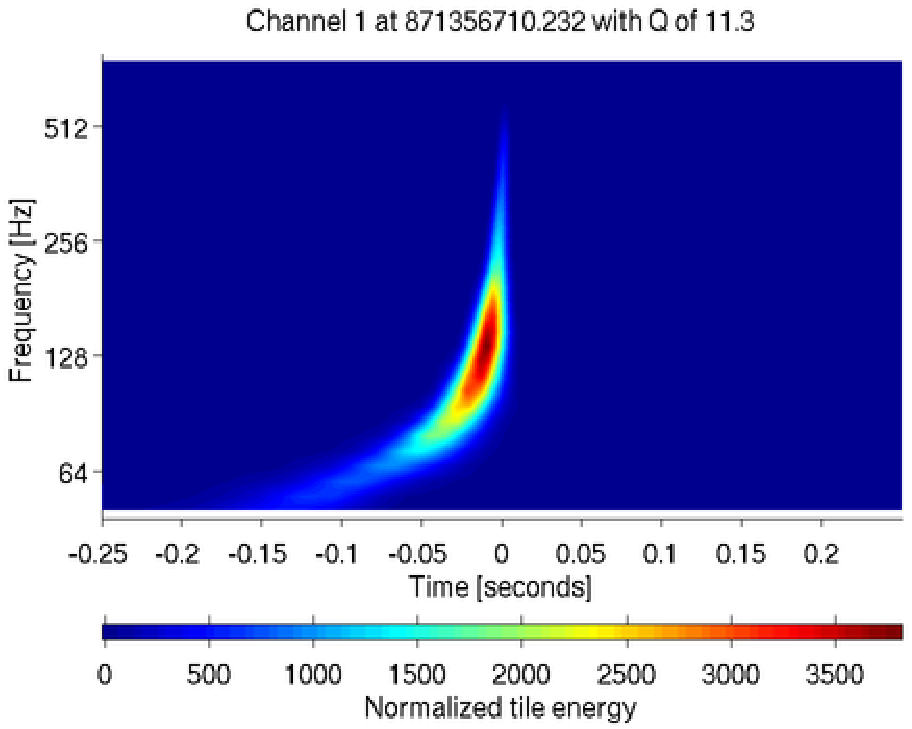}\\
\includegraphics[width=.45\textwidth]{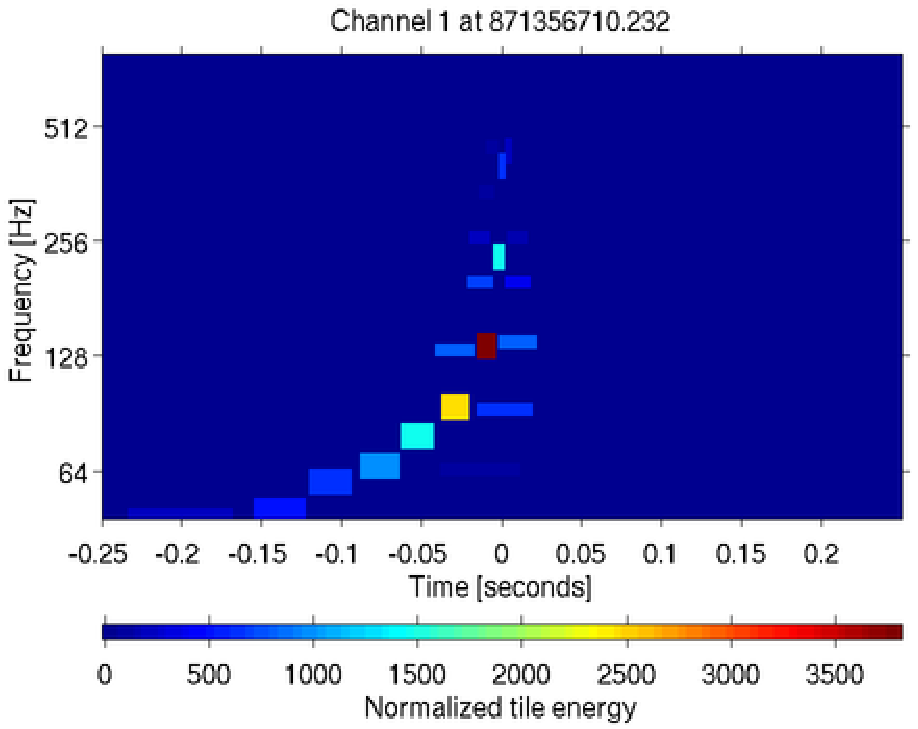}&
\includegraphics[width=.45\textwidth]{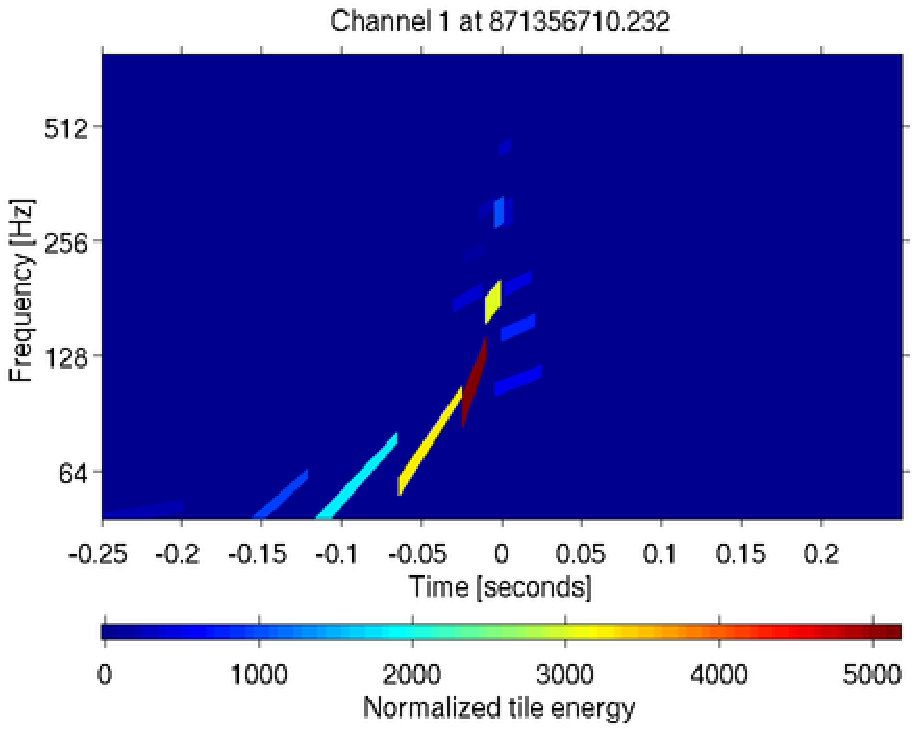}
\end{tabular}
\caption{\label{scan}(\textit{top/left}) Inspiralling black-hole binary (with
  masses $m_1=14 M_{\odot}$, $m_2=16 M_{\odot}$ and non-precessing spin
  parameters $\chi_1= -0.68$ and $\chi_2=-0.48$) signal in simulated Gaussian
  LIGO/Virgo-like noise.  (\textit{top/right}) Spectrogram
  (\textit{bottom/left}) Significant time-frequency tiles for standard Omega
  (using sine-Gaussian wavelets only) (\textit{bottom/right}) Significant
  time-frequency tiles for chirpletized Omega (using chirplets).}
\end{figure}

In Fig.~\ref{scan}, we compare the results from the standard and chirpletized
Omega pipelines obtained for a black-hole binary chirp embedded in simulated
Gaussian noise at large SNR. Chirplets with a positive slope are preferred to
sine-Gaussian with constant frequency: the correlation of the most significative
chirplet is, in this example, $\sim 45 \%$ larger than the most significative
sine-Gaussian. Work is currently in progress to understand how the background in
chirpletized Omega is different from standard Omega. Preliminary studies in
Gaussian noise suggest that the background rates are comparable, so that we can
expect an increase of $\sim 30-40 \%$ in distance reach by using chirplets.

Note that the chirplet slope provides indication of the frequency evolution of
the observed signal and thus may be very useful in the \textit{a posteriori}
interpretation of an event.

\paragraph{Systematic study ---}

We also performed a more systematic comparison over a population of inspiralling
binaries. We considered a total of 5500 binaries with equal
mass components. The total mass $M$ is extracted from a flat distribution
in the  $4-100 \, M_{\odot}$ range. The signal amplitudes are scaled so
that the SNR is distributed over an interval ranging from $\sim 10$ to
$\sim 10^3$.

\begin{figure}

\begin{center}
\includegraphics[width=.6\textwidth]{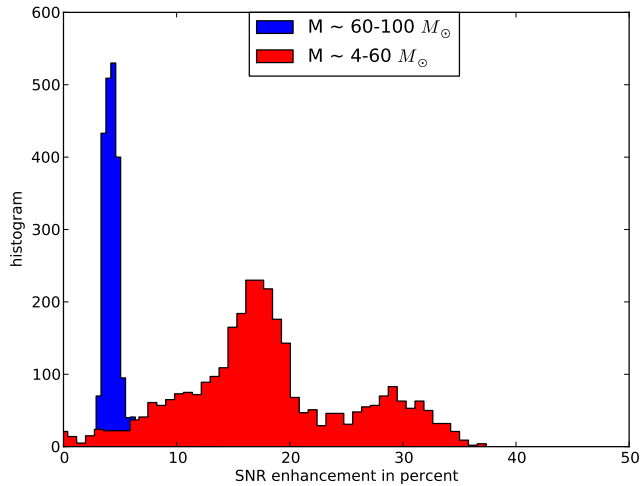}
\caption{\label{measure} SNR enhancement from standard Omega to chirpletized
  Omega. The largest improvements about $\sim 30$ to $40\%$ are mainly due to
  binaries in the low mass range $ M < 20 M_{\odot}$.}
\end{center}
\end{figure}

We obtain an estimate of the injected SNR from the amplitude of the most
significant template. In the ideal case where signal and template are identical,
the estimate equals the injected value. In Fig.~\ref{measure}, we show the
relative difference of the SNR estimated by chirpletized Omega and
standard Omega.  Chirpletized Omega estimates a higher SNR across the mass
range. However, there are two regimes: for the high-mass range $M
\gtrsim 60 M_{\odot}$, the SNR improvement is small ($\sim 5 \%$) while it is
more pronounced ($\sim 20 \%$) for the low-mass range $M \lesssim 60
M_{\odot}$. The improvement may go upto $\sim 40\%$ for  $M \lesssim 20 M_{\odot}$.

Generally speaking, the spectrum of the GW chirp is shifted toward low
frequencies when the binary mass increases. The frequency associated with the
innermost stable circular orbit (ISCO), which corresponds to the transition
between inspiral and merger phase of the coalescence, is below 70 Hz for masses
$M \gtrsim 60 M_{\odot}$.  In this condition, the chirp phase of the waveform
with a spectral content at frequencies below ISCO is outside the detector
sensitive band\footnote{This statement is valid for the LIGO detector noise
  curve which was used for the systematic study. The mass cut-off is higher for
  the Virgo detector as its sensitive band extends to lower frequencies.} and
thus does not contribute significantly to the SNR. Sine-Gaussian waveform
provides a good enough fit of the remaining few waveform cycles associated with
the merger and ringdown parts of the coalescence. This explains the two regimes
in Fig. \ref{measure}.

\section{Status and future plans}

We introduced a new chirplet-based extension of the Omega pipeline. We show
preliminary results using coalescing binary mergers waveforms. Versatility
(robustness to signal model uncertainty) and algorithmic simplicity are two
advantages of this methodology when compared to more standard approaches for the
detection of such waveforms.

The single-detector network search code is ready and it can be downloaded
\cite{omega_chirplet} and used to produce chirpletized Omega scans similar
to the one we show in Fig. \ref{scan}.

We continue to study the response of the code to real noise and we aim at a
complete, operating pipeline using chirplets as templates, and new clustering strategies
tailored to these templates, as well as a multi-detector network strategy.

\section*{References}
\bibliographystyle{plain}
\bibliography{paper}

\section*{Acknowlegments}

We are grateful to the LIGO Scientific Collaboration for the use of its
algorithm library available at
\url{https://www.lsc-group.phys.uwm.edu/daswg/projects/lalapps.html}.  This work
is supported in part by NSF grant PHY-0653550.  M. Miele is supported by an ERASMUS
fellowhip from the University of Sannio at Benevento (Italy).
\end{document}